\documentstyle[12pt]{article}
\begin{document}
\newpage
\centerline{Preprint IBR-TH-99-S-05, Jan. 3, 1999}
\centerline{in press at {\it Algebras, Groups and Geometries} {\bf 15},
473-498, 1998}
\vspace*{0.50cm}
\centerline{{\it \bf Dedicated to Marius Sophus Lie in the Centennial of
His Death}}

\vspace*{0.50cm}
\begin{center}
{\bf ISOTOPIC, GENOTOPIC AND HYPERSTRUCTURAL LIFTINGS OF LIE'S THEORY
AND
THEIR ISODUALS}
\end{center}
\centerline{{\bf Ruggero Maria Santilli}}
\centerline{Institute for Basic Research}
\centerline{P.O.Box 1577, Palm Harbor, FL 34682, U.S.A.}
\centerline{ibr@gte.net; http://home1.gte.net/ibr}

\centerline{{\bf Abstract}}

\noindent {\small After reviewing the basic role of Lie's theory for the
mathematics and physics of this century, we identify its limitations for
the treatment of systems beyond the local-differential, Hamiltonian and
canonical-unitary conditions of the original conception. We therefore
outline three sequential generalized mathematics introduced by the
author
under the name of {\it iso-, geno- and hyper-mathematics} which are
based
on generalized, Hermitean,
non-Hermitean and multi-valued units, respectively. The resulting {\it
iso-,
geno- and hyper-Lie theories}, for which the new
mathematics were submitted, have been extensively used for the
description
of nonlocal-integral systems with action-at-a-distance Hamiltonian and
short-range-contact non-Hamiltonian interactions in reversible,
irreversible and multi-valued conditions, respectively. We then point
out
that conventional, iso-, geno- and hyper-Lie theories are unable to
provide a consistent {\it classical} representation of {\it antimatter}
yielding the correct charge conjugate states at the operator
counterpart. We
therefore outline yet novel mathematics proposed by the author under the
names of {\it isodual conventional, iso-, geno- and hyper-mathematics},
which constitute anti-isomorphic images of the original mathematics
characterized by {\it negative-definite units and norms}. The emerging
isodual generalizations of Lie's theory have permitted a novel
consistent
characterization of antimatter at all levels of study, from Newton to
second quantization. The main message emerging after three decades of
investigations is that the sole generalized theories as invariant as the
original formulation, the sole usable in physics, are those preserving
the
original abstract Lie axioms, and merely realizing them in broader
forms.}

\newpage

\noindent {\large \bf 1. Majestic Consistency of
Lie's Theory}.

\noindent Since I was first exposed to the theory of {\it Marius
Sophus Lie} [1] during my graduate studies in physics at the University
of
Torino, Italy, in the 1960's, I understood that Lie's theory has a
fundamental character for the virtual entire contemporary mathematics
and
physics.

I therefore dedicated my research life to identify the {\it limitations}
of Lie's theory and construct possible {\it generalizations} for
physical conditions broader than those of the original
conception. In this paper I outline the most salient aspects of this
scientific journey (as representative references, see my original papers
in
the field [3],  mathematical studies [4],
physical studies [5], monographs [6], applications and experimental
verifications [7-10]).

Let $F = F(a,+,\times)$ be a field
of conventional numbers a (real, complex or quaternionic numbers) with
conventional sum +, (associative) product $\times$) additive unit 0 and
multiplicative unit I. When formulated on a Hilbert space $\cal H$ over
F,
the physically most
important formulation of Lie's theory is that via connected
transformations of an operator A on $\cal H$ over F in the  following
finite
and infinitesimal forms and interconnecting conjugation
$$
A(w) = U\times A(0)\times U^{\dagger} =
e^{iX\times w}\times A(0)\times e^{-iw\times X} ,
\eqno (1.1a)
$$
$$
idA / dw = A\times X - X\times A = [A, X]_{operator},
\eqno (1.1b)
$$
$$
e^{iX\times w} = [ e^{- i w\times X} ]^{\dagger},  X = X^{\dagger}, w
\epsilon F,
\eqno (1.1c)
$$
with classical counterpart in terms of vector-fields on the cotangent
bundle (phase space) with local chart $(r^k, p_k)$, k = 1, 2, 3, over F
$$
A(w) = U\times A(0)\times U^t =
e^{-w\times (\partial X/\partial r^k)\times (\partial/\partial p_k)}
\times A(0) \times
e^{w(\partial/\partial r^k)\times (\partial X/\partial p_k)},
\eqno (1.2a)
$$
$$
{dA\over dw} = {\partial A\over \partial r^k}\times {\partial X\over
\partial p_k} - {\partial X\over \partial r^k}\times {\partial A\over
\partial p_k}
 =  [A, X]_{classical},
\eqno (1.2b)
$$
and unique interconnecting map given by the conventional or
symplectic quantization.

As it is well known, Lie's theory is at the foundation of the
mathematics of this century, including
topology, vector and metric spaces, functional analysis, differential
equations, algebras and groups, geometries, etc.

As it is also well know, Lie's theory is at the foundation of all
physical
theories of this century, including classical and quantum mechanics,
particle physics, nuclear physics, superconductivity, chemistry,
astrophysics, etc. In fact, whenever the parameter w represents time t,
Eqs. (1.1) are the celebrated Heisenberg equations of motion in finite
and infinitesimal form, while Eq.s (1.2) are the classical Hamilton
equations, also in their finite and infinitesimal forms.
Characterization
via Lie's theory of all classical and operator branches of physics then
follows.

 A reason
for the majestic consistency of Lie's theory most important for physical
applications is that of being {\it form invariant} under the
transformations of its own class. In fact, connected Lie groups (1.1a)
constitute {\it unitary transforms} on
$\cal H$ over F,
$$
U\times U^{\dagger} =
U^{\dagger}\times U = I,
\eqno (1.3)
$$
under which we have the following invariance laws for units, products
and
eigenvalue equations
$$
I\rightarrow U\times I\times U^{\dagger} = I' = I,
\eqno (1.4a)
$$
$$
A\times B\rightarrow U\times (A\times B)\times U^{\dagger} =
(U\times A\times U^{\dagger})\times (U\times B\times U^{\dagger}) =
A'\times B',
\eqno (1.4b)
$$
$$
H\times |\psi > = E\times |\psi>\rightarrow U\times H\times |\psi > =
(U\times H\times U^{\dagger})\times(U\times |\psi >)
= H'\times |{\psi}'> =
$$
$$
U\times E\times |\psi> = E'\times |{\psi}'>, E' = E.
\eqno (1.4c)
$$
with corresponding invariances at the classical level here omitted for
brevity.

 It then follows that {\it Lie's theory possesses
numerically invariant units, products and
eigenvalues,} thus verifying the necessary condition for
physically consistent applications.

\vskip 0.50cm

\noindent {\large \bf 2. Initial Proposals of Generalized Theories.}

\noindent Despite the above majestic mathematical and physical
consistency, by no means Lie's theory can represent the totality of
systems existing in the universe. In fact, inspection of structures
(1.1)
and (1.2) reveals that, in its conventional formulation, {\it Lie's
theory
can only represent isolated-conservative-reversible systems of
point-like
particles with only potential-Hamiltonian internal interactions}. In
fact,
the point-like structure is demanded by the local-differential character
of
the underlying topology; the isolated-conservative character of the
systems
is established by the fact that the brackets [A, B] of the time
evolution
are totally antisymmetric, thus implying conservation laws of total
quantities; the sole potential character is established by the
representation of systems solely via a Hamiltonian; and  the
reversibility
is established by the fact that all known action-at-a-distance
interactions
are reversible in time (i.e., their time reversal image is as physical
as
the original one, as it is the case for the orbit of a planet). All
admissible interactions are represented via time-independent potentials
V in the Hamiltonian H =
$p^2/2m + V$, resulting in manifestly reversible systems.

I therefore initiated
a long term research program aiming at generalizations (I called {\it
liftings}) of Lie's theory suitable for the representation of broader
systems.

The first lifting I proposed as part of my Ph.D. thesis [3a,3b] back in
1967
is that for the representation of {\it open-nonconservative systems},
that
is, systems whose total energy H is not  conserves in time, $idH/dt \not
=
0$, because of interactions with the rest of the universe. This called
for
the formulation of the  theory in such a way that its brackets are
not totally antisymmetric. In this way I
proposed, apparently for the first time in 1967, the broader
(p-q)-parametric deformations (known in more recent times as the
q-deformations),
$$
A(w) = U\times A(0)\times U^{\dagger}
= e^{iw\times  p\times X}\times
A(0)\times e^{-iw\times q\times X}, X = X^{\dagger},
\eqno (2.1a)
$$
$$
idA / dw = p\times A\times X - q\times X\times A = (A,X)_{operator},
\eqno (2.1b)
$$
where p, q and $p+/- q$ are non-null parameters, with classical
counterpart
[3c]
$$
A(w) = U\times A(0)\times U^t = e^{-w\times q\times  (\partial
X/\partial
r^k)\times (\partial/\partial p_k)}
\times A(0) \times
e^{w\times p\times (\partial/\partial r^k)\times (\partial X/\partial
p_k)},
\eqno (2.2a)
$$
$$
{dA\over dw} = p\times {\partial A\over \partial r^k}\times {\partial
X\over
\partial p_k} - q\times {\partial X\over \partial r^k}\times {\partial
A\over
\partial p_k}
 =  (A, X)_{classical}.
\eqno (2.2b)
$$

Prior to releasing papers [3a] for publication, I spent about one year
in European mathematical libraries to identify the algebras
characterized by brackets (A, B) which resulted to be {\it
Lie-admissible} according to Albert [2] (a generally nonassociative
algebra with product (A, B)  is said to be Lie-admissible when the
attached
algebra with antisymmetric product [A, B] = (A, B) - (B, A) is Lie). At
the
time of proposal [3a] only {\it three}  papers had appeared
in Lie-admissible algebras and only in the mathematical literature (see
Ref. [3a]).

The (p, q)-parametric deformations (2.1), (2.2) did indeed achieve the
desired objective. In fact, the total energy and other physical
quantities
{\it are not} conserved by assumption, because $idH/dt = (p -
q)\times H\times H \not = 0$.

In 1968 I emigrated with my family to the U. S. A, where I soon
discovered
that Lie-admissible theories were excessively ahead of their time
because
unknown in {\it mathematical}, let alone physical circles.
Therefore, for a number of years I had to dedicated myself to more
mundane
research along the preferred lines of the time.

When I passed to Harvard University in 1978 I resumed research on
Lie-admissibility and proposed the most general possible (P,
Q)-operator Lie-admissible theory according to the operator structures
[3d]
$$
A(w) = U\times A(0)\times U^{\dagger}
= e^{iw\times X\times Q}\times
A(0)\times e^{-iw\times P\times X}, X = X^{\dagger}, P = Q^{\dagger},
\eqno (2.3a)
$$
$$
idA / dw = A\times P\times X - X\times Q\times A =
(A\hat {,} X)_{operator},
\eqno (2.3b)
$$
where $P, Q, and P+/- Q$ are non-singular matrices (or operators) such
that
P-Q characterizes Lie brackets, with classical counterpart [3d]
$$
A(w) = U\times A(0)\times U^t = e^{-w\times \times  (\partial X/\partial
r^i)\times Q^i_j\times (\partial/\partial p_k)}
\times A(0) \times
e^{w\times (\partial/\partial r^i)\times P^i_j (\partial X/\partial
p_j)},
\eqno (2.4a)
$$
$$
{dA\over dw} = {\partial A\over \partial r^i}\times P^i_j\times
{\partial
X\over
\partial p_j} - {\partial X\over \partial r^i}\times Q^i_j\times
{\partial
A\over
\partial p_j}
 =  (A, X)_{classical}.
\eqno (2.4b)
$$

A primary motivation for generalizations (2.3) and (2.4) over (2.1) and
(2.2) is that the latter constitute nonunitary-noncanonical transforms.
The application of a nonunitary transform to Eqs. (2.1) then yields
precisely Eqs. (2.3) with $P = p\times (U\times U^{\dagger})^{-1}$ and
$Q =
q\times (U\times U^{\dagger})^{-1}$, as we shall see better below. The
application of any additional nonunitary transform then preserves the
Lie-admissible structure. A similar case occurs for the classical
counterpart.

Additional studies established that structures (2.3) constitute the most
general possible transformations admitting an algebra in the
infinitesimal
form. In particular, the product $(A\hat {,} B)$ results to be jointly
{\it
Lie- and Jordan admissible}, although the attached Lie and Jordan
algebras
are more general than the conversional forms.

The latter generalized character permitted me to propose a
particularization of the above Lie-admissible theory I called {\it
Lie-isotopic} [3d,3r] , in which the brackets did verify the Lie axioms,
but
are more general then the conventional versions, with operator
formulation
$$
A(w) = U\times A(0)\times U^{\dagger} = e^{iX\times T\times w}\times
A(0)\times e^{-iw\times T\times X}, \hat T = hat T^{\dagger},
\eqno (2.5a)
$$
$$
idA / dw = A\times T\times X - X\times T\times A = [A\hat
{,}X]_{operator},
\eqno (2.5b)
$$
and classical counterpart [3d,3r]]
$$
A(w) = e^{- w\times (\partial
X/\partial r^{i})\times T^i_j\times (\partial/\partial p_j}) >
A(0)  < e^{-w(\partial/\partial p_j)\times T^i_j\times (\partial
X/\partial
r^i)},
\eqno (2.6a)
$$
$$
{dA\over dw} = {\partial A\over \partial r^i}\times T^i_j\times
{\partial
X\over
\partial p_j} - {\partial X\over \partial r^i}\times T^i_j\times
{\partial
A\over
\partial p_j}
 =  [A\hat {,} X]_{classical}.
\eqno (2.6b)
$$

As one can see, the latter theories too are nonunitary-noncanonical, and
the application of additional nonunitary-noncanonical transforms
preserves the Lie-isotopic character. This establishes that
transformations
(2.5), (2.6)  are the most general possible ones admitting a Lie algebra
in
the brackets of their infinitesimal versions.

\vskip 0.50 cm

\noindent {\large \bf 3. Inconsistencies of Initial Generalizations.}

\noindent Following the proposals of theories (2.1)-(2.6), I discovered
that, even though {\it mathematically} intriguing and significant, the
above Lie-isotopic and Lie-admissible theories had {\it no physical
applications}. This is due to the fact that all the broader theories
considered have a {\it nonunitary structure} at the operator level with
a
{\it noncanonical structure} st the classical counterpart.

In the transition from unitary to nonunitary theories, invariances
(1.4) are turned  into the following noninvariances,
$$
U\times U^{\dagger} =
U^{\dagger}\times U \not = I,
\eqno (3.1a)
$$
$$
I\rightarrow U\times I\times U^{\dagger} = I' \not = I,
\eqno (3.1b)
$$
$$
A\times B\rightarrow U\times (A\times B)\times U^{\dagger} =
$$
$$
 (U\times A\times U^{\dagger})\times (U\times
U^{\dagger})^{-1}\times (U\times B\times U^{\dagger}) =
A'\times T\times B', T = (U\times U^{\dagger})^{-1},
\eqno (3.1c)
$$
$$
H\times |\psi > = E\times |\psi>\rightarrow U\times H\times |\psi > =
(U\times H\times U^{\dagger})\times (U\times U^{\dagger})^{-1}\times (U
\times |\psi >)
=
$$
$$
H'\times T\times |{\psi}'> =
U\times E\times |\psi> = E'\times |{\psi}'>, E' \not = E,
\eqno (3.1d)
$$

It then follows that all theories with a nonunitary structure have the
following physical inconsistencies studied in detail in Refs. [12]: 1)
nonunitary theories do not have invariant units of time, space, energy,
etc., thus lacking any physically meaningful applications to
measurements
(for which the invariance of the basic units is a necessary
pre-requisite);
2) nonunitary theories do not preserve in time the original Hermiticity
of
operators, thus having no physically acceptable observables; 3)
nonunitary
theories do not have invariant conventional and special functions and
transforms, thus lacking unique and invariant numerical predictions;
nonunitary theories violate probability and causality laws; nonunitary
theories are incompatible with Galilei's and  Einstein's relativities;
and
suffer from other serious shortcomings.  Similar inconsistencies
exist at the classical level.

Corresponding mathematical inconsistencies also occur [12f,12g]. In
fact,
nonunitary theories are generally formulated on a conventional metric or
Hilbert space defined over a given field which, in turn, is based on a
given unit I. But the fundamental unit is not left invariant by
nonunitary
transforms by conception. It them follows that the entire mathematical
structure of nonunitary theories becomes inapplicable for any value of
the
parameters different than the initial values.

It should be noted that the above catastrophic inconsistencies
also hold for any other theory departing from Lie's theory, yet
formulated
via conventional mathematics, such as deformations, Kac-Moody algebras,
superalgebras, etc. [12].

After systematic studies I realized that the {\it only}
possibility to reach invariant formulations of generalized Lie theories
 was that of {\it constructing new
mathematics} specifically conceived for the task at hand.

Since no other mathematics was available for the representation of the
broader theories here considered, as a physics I had
to initiate long and laborious
mathematical studies in constructing the new mathematics, as a
pre-requisite
for conducting physical research.

Predictably, the task resulted to be more difficult than I
suspected. In fact, after having lifted all the essential aspects of
conventional mathematics (such as numbers and fields, vector and metric
spaces, algebras and groups, geometries, etc.) [3s] into the
needed broader form, I continued to miss the crucial invariance.

Insidiously, the problem resulted to exist where I was expect it the
least,
in the {\it ordinary differential calculus}. It was only in memoir [3i]
of
1966 that I finally achieved invariance following suitable liftings of
the
ordinary differential calculus. The reader is therefore warned that {\it
all papers on Lie-isotopic and Lie-admissible theories prior to memoir
[3i] have no consistent physical applications because they lack
invariance}.

The invariant liftings of Lie's theory which resulted from these
efforts can be summarized as follows.
\vskip 0.50 cm

\noindent {\large \bf 4. Lie-Santilli Isotheory.}

\noindent The main idea [3d] is the
lifting the conventional, trivial, n-dimensional unit I = diag. (1, 1,
..., 1) of Lie's theory into a real-values, nowhere singular and
positive-definite
$n\times n$-dimensional matrix $\hat I$, called {\it isounit} (where the
prefix "iso-" means "axiom-preserving"), with an unrestricted functional
dependence on time t, coordinates $r = (r^k)$, momenta p = $(p_k)$, k =
1,
2, 3, wavefunctions
$\psi$,  and any other needed variable,
$$
I= diag.(1, 1, ..., 1)\rightarrow \hat I(t, r, p, \psi, ...) = 1/\hat T
\not = I.
\eqno (4.1)
$$

The applicable mathematics, called {\it isomathematics}, is the lifting
of the {\it totality} of conventional mathematics (with a well defined
unit), without any exception known to me, into a new form admitting
$\hat
I$, rather than I, as the correct left and right unit. This calls for:

1) The lifting of the associative product $A\times B$ amon g generic
quantities A, B (such as numbers, vector-fields, operators, etc.) into
the form, called {\it isoassociative product}, for which $\hat I$ is
indeed the left and right unit,
$$
A\times B\rightarrow A\hat {\times} B = A\times \hat T\times B,
\hat I\hat {\times} A = A\times \hat {\times}\hat I = A;
\eqno (4.2)
$$

2) The lifting of fields $F = F(a,+,\times)$ into the
{\it isofields}
$\hat F = \hat F(\hat a,\hat +.\hat {\times})$ of {\it isonumbers} $\hat
a =
a\times
\hat I$ (isoreal, isocomplex and isooctonionic numbers) with {\it
isosum}
$\hat a \hat + \hat b = (a + b)\times \hat I$, {\it isoproduct} $\hat
a\hat
{\times}\hat b = (a\times b)\times \hat I$, {\it isoquotient} $\hat a
\hat
{/}\hat b = (\hat a / \hat b)\times hat I$, etc. (see [3h] for details)
;

3) The lifting of functions f(r) on F into {\it isofunctions} $\hat
f(\hat
r)$ on $\hat F$,  such as the {\it isoexponentiation} $\hat e^{\hat A} =
\hat I + \hat A/1! + \hat A\hat {\times}\hat A/2! + ... = (e^{\hat
A\times
\hat T)}\times \hat I = \hat I\times (e^{\hat T\times \hat A}$, and
related
lifting of transforms into {\it isotransforms} (see [3i,3s] for
details);

4)  The lifting of the ordinary differential calculus into the {\it
isodifferential calculus}, with basic rules
$\hat d\hat r^k =
\hat I^k_i\times d\hat r^i, \hat d\hat p_k = \hat T_k^i\times d\hat p_i$
(because
$r^k$ and $p_k$ are defined on isospaces with isometrics inverse of each
other), {\it isoderivatives}
$\hat {\partial}/\hat {\partial} \hat r^i = \hat T_i^j\times \partial
/\partial \hat r^j$, $\hat {\partial}/\hat {\partial} \hat p_k = \hat
I_k^i\times
\partial/\partial \hat p_i$, $\hat {\partial} \hat r^i/\hat {\partial}
\hat
r^j =
\hat {\delta}^i_j = {\delta}^i_j\times \hat I$, etc.(see [3i] for
details);

5) The lifting of conventional vector, metric and Hilbert spaces into
their isotopic images, e.g., the lifting of the Euclidean space
$E(r,\delta,R)$ with local coordinates $r = (r^k)$ and metric $\delta =
Diag. (1, 1,1)$ into the {\it isoeuclidean spaces} $\hat E(\hat r,\hat
{\delta},\hat R)$ with isocoordinates $\hat r = r\times \hat I$ and
isometric
$\hat {\delta} = \hat T\times \delta$ over the isoreals $\hat R$, or the
lifting of the Hilbert space $\cal H$ with inner product $<|\times
|>\times
I$ over the complex field $ C$ into the {\it isohilbert space} $\hat
{\cal H}$ with {\it isoinner product} $<|\hat {\times}|>\times \hat I$
over
the isocomplex field $\hat C$; etc.   (see [3s] for details).

6) The lifting of geometries and topologies into their corresponding
isotopic images (see [3n] for details);

7) The isotopic lifting of all various branches of Lie's theory, such as
the liftings of: universal enveloping associative algebras
(including the Poincar\'e-Birkhoff-Witt theorem), Lie's algebras
(including
Lie's first, second and third theorems); Lie's groups, transformation
and representation theory, etc.

The main operator formulation of the Lie-Santilli isotheory can
be written
$$
\hat A(\hat w) = \hat e^{i\hat X\hat {\times}\hat w}\hat {\times} \hat
A(\hat 0)\hat {\times}\hat e^{-i\hat w\hat {\times}\hat X} =
$$
$$
[e^{iw\times
\hat X\times
\hat T}\times A(0)\times e^{-iw\times \hat T\times hat X}]\times \hat I,
X = X^{\hat {\dagger}}, \hat T = \hat T^{\hat {\dagger}},
\eqno (4.3a)
$$
$$
i \hat d\hat A\hat {/}\hat d\hat w = \hat A\hat {\times} \hat X - \hat
X\hat {\times}\hat A =\hat A\times \hat T\times \hat X - \hat X\times
\hat
T\times \hat A = [A\hat {,}X]_{operator},
\eqno (4.3b)
$$
$$
hat e{i\hat X\times \hat w} = (\hat e^{-i\hat w\hat {\times}\hat
X})^{\hat
dagger},
\eqno (4.3c)
$$
with classical counterpart
$$
\hat A(\hat w) = e^{-w\times (\hat {\partial}\hat  X\hat {/}\hat
{\partial}\hat r^k)\hat \times (\hat {\partial}\hat {/}\hat
{\partial}\hat
p_k)}
\hat {\times}\hat A(\hat 0) \hat {\times}
\hat e^{w(\hat {\partial}\hat {/}\hat {\partial}\hat r^k)\hat {\times}
(\hat
{\partial}\hat X\hat {/}\hat {\partial}\hat p_k)},
\eqno (4.4a)
$$
$$
{\hat d\hat A\over \hat d\hat w} = {\hat {\partial}\hat A\over \hat
{\partial}\hat r^k}\hat {\times} {\hat {\partial} X\over
\hat {\partial}\hat p_k} - {\hat {\partial}\hat X\over \hat
{\partial}\hat
r^k}\hat {\times} {\hat {\partial}\hat A\over
\hat {\partial}\hat p_k} =
$$
$$
[ {\partial A\over \partial r^k}\times {\partial X\over
\partial p_k} - {\partial X\over \partial r^k}\times {\partial A\over
\partial p_k}]\times \hat I
 =  [A\hat {,} X]_{classical},
\eqno (4.4b)
$$
and unique interconnecting map called {\it isosymplectic quantization}
[3s].

A most salient feature of the Lie-Santilli isotheory is that it is {\it
form invariant under all possible nonunitary transforms}, thus achieving
the fundamental physical objective indicated earlier. In fact, an
arbitrary nonunitary transform on $\cal H$ over F can always be uniquely
written as the {\it isounitary transform} on $\hat {\cal H}$ over $\hat
F$,
$$
V\times V^{\dagger} = \hat I \not = I, V = \hat V\times \hat T^{1/2},
V\times V^{\dagger} = \hat V\hat {\times} \hat V^{\hat {\dagger}} =
\hat V^{\hat {\dagger}}\hat {\times} \hat V = \hat I,
\eqno (4.5)
$$
under which we have the isoinvariance laws
$$
\hat I\rightarrow \hat V\times \hat I\times \hat V^{\hat {\dagger}} =
\hat I' = \hat I,
\eqno (4.6a)
$$
$$
\hat A\hat {\times} \hat B\rightarrow \hat V\hat {\times}
(\hat A\hat {\times} \hat  B)\hat \times
\hat V^{\hat {\dagger}} = (\hat V\hat {\times}\hat A\hat {\times}
V^{\dagger})\hat
{\times} (\hat V\hat {\times}\hat B\hat {\times}\hat V^{\hat {\dagger}})
=
\hat A'\hat {\times}\hat B',
\eqno (4.6b)
$$
$$
\hat H\hat {\times} |\hat {\psi} > =
\hat E\hat {\times} |\hat {\psi}>\rightarrow \hat V\hat {\times}
\hat H\hat \times |\hat {\psi} > =
\hat V\hat {\times}\hat  H\hat {\times}\hat V^{\hat {\dagger}}
\hat {\times}\hat V\hat {\times}
|\hat {\psi} > = \hat H'\hat {\times} |{\hat {\psi}}'> =
$$
$$
\hat V\hat {\times}\hat E\hat {\times} |\hat {\psi}> =
\hat E'\hat {\times}
|{\hat {\psi}}'>, \hat E' = \hat E,
\eqno (4.6c)
$$
with corresponding isoinvariances for the classical counterpart.

As one can see,  isomathematics
achieves the {\it invariance of the numerical values
of the isounit, isoproduct and isoeigenvalue}s, thus regaining the
necessary conditions for physical applications.

It is easy to prove that
{\it isohermiticity coincides with the conventional
Hermiticity}. As a result, all conventional observables of unitary
theories remain observables under isotopies. The
preservation of Hermiticity-observability in time is then ensured by
the above isoinvariances. Detailed studies conducted in Ref. [3l]
established the resolution of all inconsistencies of nonunitary
theories.

By comparing Eqs. (1.1)-(1.2) and (4.3)-(4.4) it is evident that the Lie
theory and the Lie-Santilli isotheory coincide at the abstract level by
conception and construction [3d,3i]. In fact, the latter can be
characterized by "putting a hat" to the totality of quantities and
operations of Lie's theory with no exception known to me (otherwise the
invariance is lost).

Despite this mathematical similarity, the physical implications of the
Lie-Santilli isotheory are far reaching. By recalling that Lie's
theory is at the foundation of all of physics, Eqs. (4.3) and (4.4)
have permitted a structural generalization of the fundamental dynamical
equations of classical and quantum mechanics, superconductivity and
chemistry into new disciplines called {\it isomechanics}  [3] {\it
isosuperconductivity} [7] and {\it isochemistry} [8]. These new
disciplines essentially preserve the physical content of the old
theories,
including the preservation identically of the total conserved
quantities, but add internal nonhamiltonian effects represented by the
isounit that are outside any hope of representation via
Lie's theory.

In turn, these novel effects have permitted momentous advances in
various
scientific fields, such as the first axiomatically consistent
unification
of electroweak and gravitational interactions [3k,3q].

An illustrative classical application of the Lie-Santilli isotheory is
the
representation of the structure of Jupiter when considered isolated from
the rest of the Solar system, with action-at-a-distance gravitational
and
other interactions represented with the potential V in the Hamiltonian H
and additional, internal contact-non-Hamiltonian interactions
represented
via the isounit $\hat I$.

An illustrative operator application is
given by novel structure models of the strongly interacting particles
(called {\it hadrons}) for which the theory was constructed [3j]. In
turn,
the latter application has far reaching implications, including the
prediction of novel, clean {\it subnuclear} energies.
\vskip 0.50 cm

\noindent {\large \bf 5. Lie-Santilli Genotheory}.

\noindent  The main insufficiency of the
Lie-Santilli isotheory is that it preserves the totally antisymmetric
character of the classical and operators Lie brackets, thus being
unsuited
for a representation of open-nonconservative systems. In particular,
despite the broadening of unitary-canonical theories into
nonunitary-noncanonical extensions, the fundamental problem of the {\it
origin
of the irreversibility} of our macroscopic reality does not admit
quantitative treatment via the Lie-Santilli
isotheory because the latter theory is also is structurally reversible
(that
is, the theory coincides with its time reversal image for reversible
Hamiltonians and isounits).

The resolution of this insufficiency required the broadening
of the Lie-Santilli isotheory into a form whose brackets are
neither totally antisymmetric nor totally symmetric. In turn, the
achievement of an invariant formulation of the latter theory requires
the
construction of a new mathematics I suggested back in 1978
[3d] under the name of {\it genomathematics} (where the prefix "geno"
now
stands for "axiom-inducing").

The main idea of
genomathematics   is the selection of {\it two different generalized
units} called {\it genounits}, the first $\hat I^>$ for the {\it ordered
multiplication to the right} $A>B$, called {\it forward genoproduct},
and
the second
$^<\hat I$ for the {\it ordered multiplication to the left} $A<B$,
called {\it
backward genoproduct}, according to the general rules [3d,3i,3l]
$$
\hat I^> = 1/\hat S, A>B = A\times \hat
S\times B,
 \hat I^{>}>A = A>\hat I^> = A,
\eqno (5.1a)
$$
$$
^<\hat I = 1/\hat R, A<B = A\times \hat R\times B,
^<\hat I < A = A<^<\hat I = A,
\eqno (5.1b)
$$
$$
A = A^{\dagger}, B = B^{\dagger}, \hat R = \hat {S}^{\dagger}
\eqno (5.1c)
$$

The broader genomathematics is then given by:

1) The lifting of isofields $\hat F(\hat a,\hat +,\hat {\times})$ into
the
{\it forward and backward genofields} $\hat F^>(\hat a^>,\hat +^>,>)$
and
$^<\hat F(^<\hat c,^<\hat +,<)$ with {\it forward and backward
genonumbers}
$\hat a^> = a\times \hat I^>$ and $^<\hat a = ^<\hat I\times a$, and
related
operations [3h];

2) The lifting of isofunctions $\hat f(\hat r)$ on $\hat F$ into the
{\it forward
and backward genofunctions} $\hat f^>(\hat r^>)$ and $^<\hat f(^<\hat
r)$
on $\hat F^>$ and $^<\hat F$, respectiovely,  such as $\hat e_>^{\hat
X^>} = (e^{\hat X^>\times \hat R})\times \hat I^>$ and $\hat e_<^{^<\hat
X}
= ^<\hat I\times e^{\hat S\times ^<\hat X}$,with consequential
genotopies of transforms and functional analysis at large [3i,3s];

3) The lifting of the isodifferential calculus into the {\it forward and
backward genodifferential calculus} with main forward rules
$\hat d^>\hat r^{>k} =
\hat I^{>k}_i\times d\hat r^{>i}, \hat d^>\hat p^>_k = \hat
T^>_k{^i}\times
d\hat p^>_i$,
$\hat {\partial}^>/\hat {\partial}^> \hat r^{>i} = \hat S^>_i{^j}\times
\partial /\partial \hat r^{>j}$, $\hat {\partial}^>/\hat {\partial}^>
\hat
p^>_k =
\hat S^>_k{^i}\times
\partial/\partial \hat p^>_i$, $\hat {\partial}^> \hat r^{>i}/\hat
{\partial}^>
\hat r^{>j} =
\hat {\delta}^{>i}_j = {\delta}^i_j\times \hat I^>$, etc., and
corresponding backward rules easily obtainable via conjugation (see
[3i] for details);

4) The lifting of isotopologies, isogeometries,etc. into the dual
forward
and backward genotopic forms; and

5) The lifting of the Lie-Santilli isotheory into the genotheory,
including the genotopies of the various aspects, such as universal
enveloping associative algebras for ordered product to the right and to
the
left, etc. [3i,3r,3s].

The explicit realization of the Lie-Santilli genotheory  can be
expressed
via the following finite and infinitesimal forms with related
interconnection (at a fixed value of the parameter w, thus without its
ordering) [3i,3l]
$$
\hat A(\hat w) =  e_>^{i\hat X^> >\hat w} > \hat A(\hat 0) < e_<^{-i\hat
w<^<\hat X} =
$$
$$
[e^{i\hat X\times \hat S\times w}\times \hat {I}^>]\times \hat S\times
\hat A(\hat 0)\times \hat R\times [^<\hat {I}\times e_<^{-iw\times
\hat R\times \hat X}],
\eqno (5.2a)
$$
$$
i \hat d\hat A/\hat {d}\hat w  = \hat A < \hat X - \hat X > \hat A
=
$$
$$
\hat A\times \hat R\times \hat X - \hat X\times {\hat
S}\times \hat A = (\hat A\hat {,}\hat X)_{operator},
\eqno (5.2b)
$$
$$
^<\hat X = (\hat X^>)^{\hat {\dagger}}, \hat R = \hat {S}^{\dagger}
\eqno (5.2c)
$$
classical counterpart [3i]
$$
\hat A(\hat w) =  \hat e_>^{-\hat X^>>\hat w} >
\hat A(\hat 0) < \hat e_<^{\hat w<^<\hat X} =
$$
$$
e^{-w\times (\hat {\partial}^>\hat  X^>/\hat {\partial}^>\hat r^{>k})
> (\hat {\partial}^>/\hat {\partial}^>\hat p^>_k)}
>\hat A(\hat 0) <
\hat e^{w(\hat {^<}{\partial}/^<\hat {\partial} ^{<k}\hat r) <
(^<\hat {\partial}^<\hat X/^<\hat {\partial} ^<_k\hat p)},
\eqno (5.3a)
$$
$$
{\hat d\hat A\over \hat d\hat w} = ^<{\hat {\partial}^<\hat A\over
^<\hat
{\partial}^{<k}\hat r}< {^<\hat {\partial} X\over
^<\hat {\partial} ^<_k\hat p} - {\hat {\partial}^>\hat X^>\over \hat
{\partial}^>\hat  r^{>k}} > {\hat {\partial}^>\hat A^>\over
\hat {\partial}^>\hat p^>_k} =
$$
$$
^<\hat I\times [ {\partial A\over \partial r^k}\times {\partial X\over
\partial p_k}] - [{\partial X\over \partial r^k}\times {\partial A\over
\partial p_k}]\times \hat I^>
 =  (A\hat {,} X)_{classical}
\eqno (5.3b)
$$
with unique interconnecting map called {\it genosymplectic quantization}
[3s].

A most important feature of the Lie-Santilli genotheory is its {\it form
invariance}. This can be seen by noting that a {\it pair} of nonunitary
transforms on $\cal H$ over $\hat C$ can always be identically rewritten
as
the {\it genounitary transforms} on genohilbert spaces over genocomplex
fields,
$$
V\times V^{\dagger} \not = 1, V = ^<\hat V\times \hat R^{1/2},
V\times V^{\dagger} = ^<\hat V < ^<\hat V^{\dagger} = ^<\hat V^{\dagger}
<
^<\hat V = ^<\hat I,
\eqno (5.4a)
$$
$$
W\times W^{\dagger} \not = 1, W = \hat W^>\times \hat S^{1/2},
W\times W^{\dagger} = \hat W^> >  \hat W^{>\dagger} = \hat W^{>\dagger}
> \hat W^> = \hat I^>,
\eqno (5.4b)
$$
under which we have indeed the following forward genoinvariance laws
[3j]
$$
\hat I^>\rightarrow \hat I'^> = \hat W^> > \hat I^> > \hat W^{>\dagger}
=
\hat I^>,
\eqno (5.5a)
$$
$$
\hat A>\hat B\rightarrow \hat W^> > (\hat A > \hat B) > \hat
W^{>\dagger} =
\hat A' > \hat B',
\eqno (5.5b)
$$
$$
\hat H^> > | > = \hat E^> > | > = E\times |>\rightarrow \hat W^> > \hat
H^>
> |> = \hat H'^> > |>' =
$$
$$
\hat W^> > \hat E^> > | > = E\times |>',
\eqno (5.5c)
$$
with corresponding rules for the backward and classical counterparts.

The above rules confirm the achievement of the {\it invariance of the
numerical values of genounits, genoproducts and genoeigenvalues}, thus
permitting physically consistent applications.

By recalling again that Lie's theory is at the foundation of all of
contemporary science, the Lie-Santilli genotheory has permitted an
additional  structural generalization of classical and quantum
isomechanics,
isosuperconductivity and isochemistry into their genotopic coverings.

Intriguingly, the product $\hat A<\hat B - \hat B>\hat A = \hat A\times
{\hat R}\times \hat B - \hat B\times {\hat S}\times \hat A, \hat R\not =
\hat S$, is manifestly non-Lie on conventional spaces over conventional
fields, yet it becomes fully
antisymmetry and Lie when formulates on the bimodule of the
respective envelopes to the left and to the right,
$\{^<\hat {\cal A},\hat {\cal A}^>\}$ (explicitly, the numerical
values of $\hat A<\hat B = \hat A\times \hat R\times \hat B$ computed
with
respect to $^<\hat I = 1/\hat R$ is the same as that of $\hat A>\hat B =
\hat A\times {\hat S}\times
\hat B$ when computed with respect to $\hat I^> = 1/\hat S$) [3i,3l].

A primary feature of the broader classical and operator genotheories is
that
it represents  open-nonconservative systems, as desired, because now the
total energy H is not conserved in our spacetime, $idH/dt  = H\times
(\hat
R -
\hat S)\times H
\not = 0$. Yet, the notion of
{\it genohermiticity} on $\hat {\cal H}^>$ over $\hat C^>$ coincides
with
conventional Hermiticity. Therefore, the Lie-admissible theory provides
the only  operator representation of open systems
known to this author in which
the {\it nonconserved Hamiltonian and other quantities are
Hermitean, thus observable}. In other treatments
of nonconservative systems the Hamiltonian is
generally {\it nonhermitean} and, therefore, {\it not observable}.

More importantly, genotheories have permitted a resolution of the
historical problem of the {\it origin of irreversibility} via its
reduction
to the ultimate possible layers of nature, such as particles in the core
of a star. The interested reader can find the invariant genotopic
formulations of: Newton's equations in Ref. [3i]; Hamilton's equations
with
external terms in Ref. [3i]; quantization for open-irreversible systems
in
Ref. [3i,3l]; operator theory of open-irreversible systems in Ref. [3l].
\vskip 0.50 cm

\noindent {\large \bf 6. Lie-Santilli Hypertheory}.

\noindent By no means
genotheories are sufficient to represent the entirely of nature, e.g.,
because they are unable to represent {\it biological
structures} such as a cell or a sea shell. The latter systems are indeed
open-nonconservative-irreversible, yet they possess a structure
dramatically more complex than that of a nonconservative Newtonian
system.
A study of the issue has revealed that the limitation of genotheories is
due
to their {\it single-valued character.}

As an illustration,
mathematical treatments complemented with computer visualization [10]
have
established that the {\it shape}
of sea shells can be well described via the conventional single-valued
three-dimensional Euclidean space and geometry according to the
empirical
perception of our
three Eustachian tubes. However, the same
space and geometry are basically insufficient to represent
{\it the growth in time} of sea shells. In fact,
computer visualization shows that, under the exact imposition of the
Euclidean axioms, sea shells first grow in time
in a distorted way and then crack.

Illert [10] showed that a
 minimally consistent representation of the sea shells growth in time
requires {\it six dimensions}. But sea shells exist in our environment
and
can be observed via our {\it three-dimensional} perception.
The solution of this apparent dichotomy I proposed [10] is that
via {\it multi-valued hypermathematics}  essentially characterized
by
the relaxation of the single-valued nature of the genounits
while preserving their nonsymmetric character (as
a necessary condition to represent
irreversible events), according to
the  rules [3i,3t]
$$
\hat I^> = \{\hat {I}_1^>, \hat {I}_2^>, \hat {I}_3^>, ...\} = 1/\hat S,
\eqno (6.1a)
$$
$$
A>B = \{ A\times \hat S_1\times B, A\times \hat S_2\times B,
A\times \hat S_3\times B, ...\},
 \hat I^{>}>A = A>\hat I^> = A\times I,
\eqno (6.1b)
$$
$$
^<\hat I = \{^<\hat {I}_1, ^<\hat {I}_2, ^<\hat {I}_3, ...\} = 1/\hat R,
\eqno (6.1c)
$$
$$
A<B = \{A\times \hat R_1\times B, A\times hat R_2\times B,
A\times \hat R_3\times B, ...\}
^<\hat I < A = A<^<\hat I = I\times A,
\eqno (6.1d)
$$
$$
A = A^{\dagger}, B = B^{\dagger}, \hat R = \hat {S}^{\dagger}.
\eqno (6.1e)
$$

All aspects of the bimodular genotheories
admit a unique, and significant extension to
the above hyperstructures and their explicit form is here omitted for
brevity [3i,3t]. The expression of the theory via {\it  weak equalities
and operations } was first studied by Santilli and Vougiouklis in Ref.
[11].
\vskip 0.50 cm

\noindent {\large \bf 7. Isodual theories.}

 \noindent Mathematicians appear to be unaware of the fact that,
contrary
to popular beliefs, {\it the totality of contemporary mathematics,
including
its isotopic, genotopic and hyperstructural liftings, cannot provide a
consistent classical representation of antimatter}. In fact, all these
mathematics admit {\it only one quantization channel}. As
a result, the operator image of any classical treatment of antimatter
via
these mathematics simply cannot yield the correct
charge conjugate state, but it merely yields a particle with the wrong
sign of the charge.

The occurrence should not be surprising because the study of antimatter
constitutes one of the biggest scientific unbalances of this century. In
fact, matter is treated at all possible mathematical and physical
levels,
from Newton's equations and underlying topology, all the way to second
quantization and quantum field theories, while antimatter is solely
treated
at the level of {\it second quantization}. However, astrophysical
evidence suggests quite strongly the existence of macroscopic amounts of
antimatter in the universe, to the point that even  entire galaxies and
quasars could eventually result to be made up entirely of antimatter.

The only possible resolution of this historical unbalance is that via
the
construction of a {\it yet new mathematics}, specifically conceived for
a
consistent {\it classical}  representation of antimatter whose operator
counterpart yields indeed the correct charge conjugate states.

Recall that charge conjugation is anti-homomorphic, although solely
applies at the operator level. It then follows that the new
mathematics for antimatter should be, more generally, anti-isomorphic
and
applicable at all levels of study.

After a laborious research, I proposed back in 1985 [3g] the {\it
isodual
mathematics}, namely, mathematics constructed via the {\it isodual map}
of
numbers, fields, spaces, algebras, geometries,
etc..

The {\it isodual conventional mathematics} is
characterized by the simplest conceivable anti-isomorphic map of the
unit into its {\it negative-definite form},
$$
I > 0\rightarrow -I = I^d < 0,
\eqno (7.1)
$$
under which we have the transformation law of a generic, scalar,
real-valued
quantity
$$
A(w)\rightarrow A^d(w^d) = -A(-w),
\eqno (7.2)
$$
with reconstruction of numbers, fields, spaces, algebras, geometries,
quantization, etc. in such a way to admit $I^d$, rather than I, as the
correct left and right unit.

The isodual map characterizing the broader {\it isodual iso-, geno- and
hyper-mathematics} is instead given by
$$
\hat I(\hat t, \hat r, \hat p, \hat {\psi}, ...)\rightarrow -\hat
I^{\dagger}(-\hat t^{\dagger},-\hat r^{\dagger}, -\hat p^{\dagger},
-\hat
{\psi}^{\dagger}, ...) = \hat I^d(\hat t^d, \hat r^d, \hat p^d, \hat
{\psi}^d, ...),
\eqno (7.3)
$$
and consequential reconstruction of the entire formalism to admit $\hat
I^d$ as the correct left and right new unit.

The above map is not trivial, e.g., because it implies
the reversal of the sign of {\it all} physical characteristics of matter
(and not only of the charge). As such, isodual theories provide a novel
intriguing representation of antimatter which {\it begins} at the
primitive
classical Newtonian level, as desired, and then persists at all
subsequent
levels, including that of second quantization, in which case isoduality
becomes equivalent to charge conjugation [3m].

The most general mathematics presented in this paper
is the {\it isoselfdual hypermathematics} [3i], namely, a
hypermathematics that coincides with its isodual, and is
evidently given by hypermathematics multiplied by its isodual. The
latter
mathematics has been used for one of the most general known cosmologies
[3p] inclusive of antimatter as well as of biological structures (as any
cosmology should be),
 in which the universe: has a
multi-valued structure perceived by our Eustachian tubes as a
single-valued
three-dimensional structure; admits equal amounts of matter and
antimatter
(in its
limit formulation verifying Lie's conjugation (1.1c)); removes any need
for the "missing mass"; reduces considerably the currently
believed dimension of the measured universe;  possesses all
{\it identically null} total characteristics of time,
energy, linear and angular
momentum, etc.; eliminates any singularity at the time of creation.
\vskip 0.50 cm

\noindent {\large \bf 8. Simple Construction of Generalized
Theories.}

\noindent Unpredictably, the need for new mathematics has been a major
obstacle for the propagation of the generalized Lie theories outlined in
this paper in both mathematical and physical circles.

I would like to indicate here that {\it all} generalized Lie theories,
{\it all} their underlying new mathematics and {\it all}  their
applications can be uniquely and unambiguously constructed via the
following elementary means accessible to undergraduate students.

First, isotheories can be constructed via the systematic
application of the following
nonunitary transform
$$
U\times U^{\dagger} = \hat I, (U\times U^{\dagger})^{-1} = \hat T,
\eqno (8.1)
$$
to the {\it totality} of the original formalism with no exceptions.

In fact, transform (8.1) yields the isonumbers $U\times n\times
U^{\dagger}
=  n\times \hat I$, the isoproduct, $U\times (A\times B)\times
U^{\dagger}
=
(U\times A\times U^{\dagger})\times (U\times U^{\dagger})^{-1}\times
(U\times B\times U^{\dagger}) = A'\times \hat T\times B' = A'\hat
{\times}
B'$; the correct isofunctions, such as $U\times e^X\times U^{\dagger} =
\hat
e^{\hat X}$; and the correct expression of all other aspects, including
the
Lie-Santilli isotheory and its underlying basic theorems.

Once the isotopic structure has been achieved in this way,
its invariance is proved via the reformulation of nonunitary
transforms in the isounitary form (4.5), with consequential invariance
of the isotheory as in Eqs. (4.6).

The construction of the Lie-Santilli genotheory is equally elementary,
and
requires the use,  this time, of {\it two} nonunitary transforms
$$
U\times U^{\dagger} \not = I, W\times W^{\dagger} \not = I,
U\times W^{\dagger} = \hat I^>, W\times U^{\dagger} = ^<\hat I,
\eqno (8.2)
$$
to the {\it totality} of the original formalism, again, without any
exceptions.

In fact, transforms (8.2) yields the correct form of forward and
backward
genonumbers, e.g.,
$U\times n\times W^{\dagger} =  n\times \hat I^>$, the correct form of
the
forward and backward genoproduct, genofunctions and genotransforms,
including the correct structure and representation of the Lie-Santilli
genotheory.  Once reached in this way, the
invariance  is  proved by rewriting the  nonunitary transforms in
their genounitary version (5.4). Genoinvariant laws (5.5) then follow.

The Lie-Santilli hypertheory can be constructed and proved to be
invariant  via the mere relaxation of the
single-valued character of the genounits. The explicit construction is
is here omitted for brevity [3t]).

Finally, the isodual Lie theory can be easily constructed via the
systematic application of the anti-isomorphic transform
$$
U\times U^{\dagger} = -I = I^d,
\eqno (8.3)
$$
to the totality of the original formalism with no exceptions.

This yields isodual numbers, fields, products, functions, etc. The
isodualities of isotopic, genotopic and hyperstructural theories can be
similarly constructed via the anti-isomorphic images of the preceding
transforms.

Note that the above methods is useful on both mathematical and physical
grounds. On mathematical grounds one can start from one given structure,
e.g., the representation of the conventional Poincar\'e symmetry and
construct explicitly all infinitely possible irreps of the
Poincar\'e-Santilli iso-, geno- and hyper-symmetries as well as their
isoduals [3,4].

The
methods is also useful for the ongoing efforts to unify all simple Lie
groups of the same dimension in Cartan's classification (over a field of
characteristic zero) into one single isogroup, whose study has been
initiated by Gr. Tsagas and his group [4].

On physical grounds, the method presented in this section is also
particularly valuable to generalize existing applications of Lie's
theory
via the appropriate selection of the nonunitary transform
representing the missing characteristics or properties, e.g., the
representation of a locally varying speed of light.
\vskip 0.5cm

\noindent {\large \bf 9. Ultimate Significance of Lie's Axioms.}

\noindent A unitary Lie group has
the structure of a {\it bi-module} in both its finite and infinitesimal
forms with an action from the left
$U^> = e^{iX\times w}$ and an action from the the right $^<U
=e^{-iw\times X}$ interconnected by Hermitean conjugation (1.1c) [3e].
Eqs.
(1.1) can then be written
$$
A(w) = U^> > A(0) < ^<U = e^{iX\times w}> A(0) < e^{-iw\times W},
\eqno (9.1a)
$$
$$
idA/dt = A < X - X > A,
\eqno(9.1b)
$$
$$
^<U = (U^>)^{\dagger}, X = X^{\dagger}.
\eqno (9.1c)
$$

In the Lie case both products $A<B$ and $A>B$ are evidently conventional
associative products, $A<B = A>B = A\times B$, resulting in
Lie's bimodule. However,  axiomatic  structure (9.1) {\it does not}
require
that such products  have necessarily to be conventionally associative,
because they can also be isoassociative, thus yielding
the Lie-Santilli isotheory.
Moreover, axioms (9.1) {\it do not} require that the forward and
backward
isoassociative products have to be necessarily the same,
because they can also be different,
provided that conjugation (9.1c) is met. In the latter case the
axioms   yield the Lie-Santilli genotheory with an easy extension to the
hypertheory via multi-valued realizations. Isodual theories emerge along
similar lines because axioms (9.1) do not necessarily demand that the
underlying unit be positive-definite.

It then follows that the axiomatic consistency and invariance of the
generalized theories studied in this paper can be inferred from the
original invariance of
 Lie's theory itself,  of course, when treated with the
the mathematics leaving invariant the basic
units. The only  applicable mathematics are then the iso-,
geno-, and hyper-mathematics and their isoduals.

In conclusion, by looking in retrospect some three decades of studies on
the topics outlined in this paper, the emerging most important
message  is that {\it the sole invariant classical and operator theories
are those preserving the abstract Lie axioms}, Eq.s (1.1) and (1.2),
{\it
and merely providing
their broder realizations treated with the appropriate mathematics.}

\end{document}